\documentclass[usenatbib,usegraphicx]{mn2e}
\usepackage[dvipdfm]{hyperref}
\usepackage{amssymb}
\usepackage{bm}

\newcommand {\be} {\begin{equation}}
\newcommand {\ee} {\end{equation}}
\newcommand {\bea} {\begin{eqnarray}}
\newcommand {\eea} {\end{eqnarray}}

\newcommand{\eg}{{\it e.g.}}

\newcommand{\pder}[2]{\ensuremath{\frac{\partial #1}{\partial #2}}}

\title[CDI with poloidal velocity shear]
{The effect of poloidal velocity shear on the local development of current-driven instabilities}
\author[Nalewajko \& Begelman]
{Krzysztof Nalewajko,$^1$\thanks{E-mail: knalew@jila.colorado.edu}
Mitchell C. Begelman,$^{1,2}$\\
$^1$JILA, University of Colorado and National Institute of Standards and Technology, 440 UCB, Boulder, CO 80309, USA\\
$^2$Department of Astrophysical and Planetary Sciences, University of Colorado, UCB 389, Boulder, CO 80309, USA}

\begin{document}

\maketitle

\begin{abstract}
We perform a local (short-wavelength) linear stability analysis of an axisymmetric column of magnetized plasma with a nearly toroidal magnetic field and a smooth poloidal velocity shear by perturbing the equations of relativistic magnetohydrodynamics. We identify two types of unstable modes, which we call `exponential' and `overstable', respectively. The exponential modes are present in the static equilibria and their growth rates decrease with increasing velocity shear. The overstable modes are driven by the effects of velocity shear and dominate the exponential modes for sufficiently high shear values. We argue that these local instabilities can provide an important energy dissipation mechanism in astrophysical relativistic jets. Strong co-moving velocity shear arises naturally in the magnetic acceleration mechanism, therefore it may play a crucial role in converting Poynting-flux-dominated jets into matter-dominated jets, regulating the global acceleration and collimation processes, and producing the observed emission of blazars and gamma-ray bursts.
\end{abstract}


\section{Introduction}

Many high-energy astrophysical phenomena are powered by jets that form in the vicinity of compact objects and are accelerated to relativistic speeds (blazars and gamma-ray bursts (GRBs)) or very high subrelativistic speeds (X-ray binaries, pulsar wind nebulae). The currently standard model of magnetic acceleration requires that jets be initially dominated strongly by their Poynting flux, and that acceleration takes a few orders of magnitude in distance scale \citep[\eg,][]{2007MNRAS.380...51K,2009ApJ...698.1570L}. In this process, the jet expands laterally far beyond the light cylinder of the central object, stretching the magnetic field into a nearly toroidal configuration. Magnetically-dominated plasma with a toroidal magnetic field is known to be highly susceptible to Current-Driven Instability (CDI). These instabilities cause severe problems in laboratory experiments on `Z-pinch' plasma systems. However, in the context of astrophysical jets they provide both a challenge and an opportunity. The challenge is not to destroy the jet as it propagates over several orders of magnitude in distance. The opportunity is to provide a mechanism of energy dissipation --- by tapping free magnetic energy and converting it to particle energy --- that can power high-energy emission from blazars and GRBs \citep{2006A&A...450..887G}.

Global, three-dimensional general-relativistic magnetohydrodynamical (3D GRMHD) numerical simulations of jets, tracing them from their formation around spinning black holes over distances of thousands of gravitational radii, indicate that they are marginally stable to long-wavelength modes of CDI, including the most dangerous pinch ($m=0$) and kink ($m=1$) modes \citep{2009MNRAS.394L.126M}. On the other hand, it is expected that the growth rate of CDI is similar to the Alfv{\'e}n crossing time scale of the jet, and thus should not be sensitive to the wavelength. Short-wavelength (local) modes are particularly important for the problem of energy dissipation, since the dissipation rate scales like the inverse square of the wavelength. These modes were predicted analytically to play an important role in AGN jets and pulsar wind nebulae \citep{1998ApJ...493..291B}, and recent local numerical simulations have confirmed their general properties \citep{2009ApJ...700..684M,2011ApJ...728...90M,2012MNRAS.422.1436O}.

Most analytical studies of CDI have been performed in the force-free approximation, i.e. neglecting the pressure and inertia of the matter content of jet plasma \citep{1996MNRAS.281....1I,1999MNRAS.308.1006L,2001PhRvD..64l3003T,2009ApJ...697.1681N}. This approximation is valid in the innermost regions of jets, within the fast magnetosonic point, at which the plasma magnetization $\sigma$ is still substantial. Beyond the fast magnetosonic point, the magnetic nozzle effect results in further gradual jet acceleration, and a decrease in the jet magnetization to the order of unity \citep{1994ApJ...426..269B}. In this jet region, full MHD force balance is necessary to study plasma stability, as was done by \cite{1998ApJ...493..291B}. However, that study assumed a static background configuration, while it is known that the efficiency of jet acceleration is strong function of cylindrical radius \citep{2007MNRAS.380...51K,2008MNRAS.388..551T}. Within the jet acceleration region, a strong shear of the poloidal velocity can be expected to arise, and this effect needs to be incorporated in the local stability analysis.

Velocity shear is the driving factor of the Kelvin-Helmholtz Instability (KHI), which has been studied extensively in the context of astrophysical jets. Most analytical studies of KHI have adopted a vortex-sheet background configuration, i.e. a parallel discontinuity of the velocity field \citep{1976MNRAS.176..421T,1976MNRAS.176..443B,1978A&A....64...43F,1984MNRAS.208..887B,2007ApJ...664...26H}. In some works, a more gradual transition region of finite width was considered \citep{1982MNRAS.198..617R,1982MNRAS.198.1065F,1991MNRAS.252..505B}. However, we have found no analytical study that deals with a strictly smooth velocity shear, which is natural for the local stability problem that we study in the context of CDI. Several numerical studies have investigated the effect of velocity shear on CDI or the interaction between CDI and KHI modes. Some of them have adopted the vortex-sheet approximation \citep{2011ApJ...734...19M}, while some use a gradual velocity shear with a well-localized region of strong shear generating typical KHI modes \citep{2002ApJ...580..800B}. For moderate shear values, the distinction between CDI and KHI modes becomes quite vague.

In this work we study local CDI modes of cylindrical MHD equilibria in the presence of strong but gradual velocity shear. In Section \ref{sec_disp}, we derive the dispersion relation for a sheared plasma column, which is given by Equation (\ref{eq_disp}). In Section \ref{sec_stab}, we explore the complex space of mode frequencies, looking for the fastest-growing mode as a function of the background equilibrium parameters. In Section \ref{sec_appl}, we discuss the relevance of our results to global models of astrophysical jets. We summarize our results in Section \ref{sec_summ}.


In this work, we use cgs units with $c=1$. In the linear analysis, all quantities describing the background configuration are denoted with subscript `0' (zero-order terms), while all quantities describing the linear perturbation are denoted with subscript `1' (linear terms). Vector quantities are denoted with a bold font.

\section{Derivation of the dispersion relation}
\label{sec_disp}

Here, we perform a rigorous local linear perturbation analysis of a cylindrically symmetric plasma column filled with a predominantly toroidal magnetic field, in equilibrium with a gas pressure gradient in the radial direction. We generalize the results of \cite{1998ApJ...493..291B}, taking into account a smooth poloidal velocity shear.

We introduce cylindrical coordinates $(r,\phi,z)$. Calculations are made at a certain radius $r$ in the local co-moving frame. Hence, the local poloidal plasma velocity is $v_{\rm 0,z}=0$, while the local poloidal velocity shear is $u=\partial_{\ln r}v_{\rm 0,z}>0$ (as we show later, the choice of the sign of $u$ is not significant). We assume no background rotation ($v_{\rm 0,\phi}=0$) or radial expansion ($v_{\rm 0,r}=0$). The toroidal magnetic field $B_{\rm 0,\phi}$ is radially dependent, with logarithmic derivative $\alpha_{\rm\phi}=\partial_{\ln r}(\ln B_{\rm 0,\phi})$. The poloidal magnetic field $B_{\rm 0,z}\ll B_{\rm 0,\phi}$ is a constant\footnote{Introducing a radial dependence of the poloidal magnetic field has no effect on the dispersion relation under the ordering of terms described later on.}. There is no background radial magnetic field ($B_{\rm 0,r}=0$). The gas pressure $p_0$ is in equilibrium with the magnetic field:
\be
\label{eq_fr}
\zeta p_0 =
-(\alpha_{\rm\phi}+1)\frac{B_{\rm 0,\phi}^2}{4\pi}
+v_{\rm 0,z}u\frac{B_{\rm 0,\phi}^2}{4\pi}\,,
\ee
where $\zeta = \partial_{\ln r}(\ln p_0)$. Here, we keep the term including $v_{\rm 0,z}$ as it will be eventually differentiated with respect to $r$.

We introduce linear velocity perturbations of the form
\be
\bm{v}_1(t,r,\phi,z)=\tilde{\bm{v}}_1\exp(i\omega t+ilr+im\phi+ikz)\,.
\ee
These induce perturbations of the gas pressure ($p_1$) and the magnetic field ($\bm{B}_1$). The pressure perturbation is found by perturbing the equation of state ${\rm d}_t(p\rho^{-\gamma})=0$ in combination with the continuity equation $\partial_t\rho+\bm\nabla\cdot(\rho\bm{v})=0$:
\be
\partial_tp_1+(\bm{v}_1\cdot\bm\nabla)p_0+(\bm{v}_0\cdot\bm\nabla)p_1+\gamma p_0(\bm\nabla\cdot\bm{v}_1) = 0\,,
\ee
the solution to which is
\be
p_1 = \frac{i(\zeta+\gamma)v_{\rm 1,r}-m\gamma v_{\rm 1,\phi}-\gamma v_{\rm 1,lrkz}}{(\omega+kv_{\rm 0,z})r}p_0\,,
\ee
where $v_{\rm 1,lrkz}=(lv_{\rm 1,r}+kv_{\rm 1,z})r$ is the perturbed velocity component that we will use instead of $v_{\rm 1,z}$. The magnetic field perturbation is found by perturbing the Maxwell equation $\bm\nabla\times\bm{E}=-\partial_t\bm{B}$ together with $\bm{E}=\bm{B}\times\bm{v}$:
\bea
\partial_t\bm{B}_1 &=&
(\bm{B}_1\cdot\bm\nabla)\bm{v}_0
+(\bm{B}_0\cdot\bm\nabla)\bm{v}_1
-(\bm{v}_0\cdot\bm\nabla)\bm{B}_1
\nonumber\\
&&
-(\bm{v}_1\cdot\bm\nabla)\bm{B}_0
-(\bm\nabla\cdot\bm{v}_1)\bm{B}_0\,.
\eea
The solution is:
\bea
B_{\rm 1,r} &=& \frac{(m+\eta)v_{\rm 1,r}}{(\omega+kv_{\rm 0,z})r}B_{\rm 0,\phi}\,,
\\
B_{\rm 1,\phi} &=&
\frac{i\alpha_{\rm\phi}v_{\rm 1,r}+(\eta v_{\rm 1,\phi}-v_{\rm 1,lrkz})}{(\omega+kv_{\rm 0,z})r}B_{\rm 0,\phi}\,,
\\
\label{eq_B1z}
B_{\rm 1,z} &=&
\frac{i[(m+\eta)ilr+\eta]v_{\rm 1,r}-m(\eta v_{\rm 1,\phi}-v_{\rm 1,lrkz})}{(\omega+kv_{\rm 0,z})kr^2}B_{\rm 0,\phi}
\nonumber\\
&&
-\frac{i(m+\eta)uv_{\rm 1,r}}{(\omega+kv_{\rm 0,z})^2r^2}B_{\rm 0,\phi}\,,
\eea
where $\eta=kr(B_{\rm 0,z}/B_{\rm 0,\phi})$. We note that $B_{\rm 1,z}$ contains a non-vanishing term proportional to the background velocity shear.

\subsection{Energy-momentum conservation law}

Energy-momentum conservation is governed by the equation $\partial_\mu T^{\mu\nu}=0$, where
\bea
T^{00} &=& \Gamma^2w-p+\frac{1}{8\pi}\left(E^2+B^2\right)\\
T^{0i} &=& \Gamma^2wv^i+\frac{1}{4\pi}\left(\bm{E}\times\bm{B}\right)^i\\
T^{ij} &=& \Gamma^2wv^iv^j+\left(p+\frac{E^2+B^2}{8\pi}\right)\delta^{ij}
\nonumber\\
&&
-\frac{1}{4\pi}\left(E^iE^j+B^iB^j\right)
\eea
is the stress tensor in the ideal relativistic MHD regime, $\Gamma=(1-v^2)^{-1/2}$ is the Lorentz factor and $w$ is the relativistic enthalpy. The combined energy-momentum equation is
\bea
\Gamma^2w\left(\partial_t+\bm{v}\cdot\bm\nabla\right)\bm{v}
+\left(\bm{v}\partial_t+\bm\nabla\right)p
\nonumber\\
-\rho_{\rm e}\bm{E}
-\bm{j}\times\bm{B}
+(\bm{E}\cdot\bm{j})\bm{v}
&=& 0\,,
\eea
where $\rho_{\rm e}$ is the electric charge density and $\bm{j}$ is the electric current density, which can be found from the Maxwell equations:
\bea
\rho_{\rm e} &=& \frac{1}{4\pi}(\bm\nabla\cdot\bm{E}) \\
\bm{j} &=& \frac{1}{4\pi}(\bm\nabla\times\bm{B}-\partial_t\bm{E})\,.
\eea
The perturbed energy-momentum equation is
\bea
\label{eq_pert_ene_mom}
\Gamma_0^2w_0\left[\partial_t\bm{v}_1+(\bm{v}_1\cdot\bm\nabla)\bm{v}_0+(\bm{v}_0\cdot\bm\nabla)\bm{v}_1\right]
\nonumber\\
+\bm{v}_0\partial_tp_1+\bm\nabla{p}_1-\rho_{\rm e,1}\bm{E}_0-\rho_{\rm e,0}\bm{E}_1
\nonumber\\
-\bm{j}_1\times\bm{B}_0-\bm{j}_0\times\bm{B}_1+(\bm{E}_1\cdot\bm{j}_0)\bm{v}_0+(\bm{E}_0\cdot\bm{j}_1)\bm{v}_0
&=& 0\,.
\eea

We restrict this analysis to perturbations of short wavelength in the $z$ direction, i.e. $kr\gg 1$. At the same time, we assume that $B_{\rm 0,z}\ll B_{\rm 0,\phi}$, so that $\eta\sim\mathcal{O}(1)$. For the wavelength in the $r$ direction, we assume only that $l\lesssim\mathcal{O}(k)$. We further assume that $m,u\sim\mathcal{O}(1)$; and introduce the dimensionless frequency $\tilde\omega=\omega r\sim\mathcal{O}(1)$, the magnetization parameter $\sigma_0=B_0^2/(4\pi w_0)\sim\mathcal{O}(1)$, and the sound speed $v_{\rm s}^2=\gamma p_0/w_0\sim\mathcal{O}(1)$. We can now calculate each term of the energy-momentum equation (see Appendix \ref{sec_app1}).

The azimuthal momentum equation leads to (keeping the leading terms only):
\bea
\left(mv_{\rm s}^2+\eta\sigma_0\right)v_{\rm 1,r}
+\left(m^2v_{\rm s}^2+\eta^2\sigma_0-\tilde\omega^2\right)iv_{\rm 1,\phi}
\nonumber\\
+\left(mv_{\rm s}^2-\eta\sigma_0\right)iv_{\rm 1,lrkz}
&=& 0\,,
\eea
while the poloidal momentum equation gives:
\bea
(v_{\rm s}^2-\sigma_0)v_{\rm 1,r}
+(mv_{\rm s}^2-\eta\sigma_0)iv_{\rm 1,\phi}
\nonumber\\
+(v_{\rm s}^2+\sigma_0)iv_{\rm 1,lrkz}
&=& 0\,.
\eea
We can combine these two equations to derive explicit formulas relating the perturbation velocity components:
\bea
\label{eq_v1phi}
v_{\rm 1,\phi}
&=&
\frac{2(m+\eta)v_{\rm s}^2\sigma_0}{(m+\eta)^2v_{\rm s}^2\sigma_0-\tilde\omega^2(v_{\rm s}^2+\sigma_0)}iv_{\rm 1,r}\,,
\\
\label{eq_v1lrkz}
v_{\rm 1,lrkz}
&=&
\frac{(\eta^2-m^2)v_{\rm s}^2\sigma_0-(v_{\rm s}^2-\sigma_0)\tilde\omega^2}{(m+\eta)^2v_{\rm s}^2\sigma_0-\tilde\omega^2(v_{\rm s}^2+\sigma_0)}iv_{\rm 1,r}\,.
\eea
This means that the components $v_{\rm 1,r},v_{\rm 1,\phi},v_{\rm 1,lrkz}$ of the perturbation velocity are of the same order. We can now compare the orders of terms involving different velocity components.

To $\mathcal{O}(v_1)$, the radial momentum equation multiplied by a factor $(r/w_0)$ is
\bea
\label{eq_Fr}
\frac{\partial_{\ln r}p_1}{w_0}
+\left[(1+\sigma_0)\tilde\omega-\frac{il}{k}\sigma_0u\right]iv_{\rm 1,r}
-i(m+\eta)\sigma_0\frac{B_{\rm 1,r}}{B_{\rm 0,\phi}}
\\
+(\alpha_{\rm\phi}+2)\sigma_0\frac{B_{\rm 1,\phi}}{B_{\rm 0,\phi}}
+\sigma_0\left(\frac{\partial_{\ln r}B_{\rm 1,\phi}}{B_{\rm 0,\phi}}+\frac{\eta}{kr}\frac{\partial_{\ln r}B_{\rm 1,z}}{B_{\rm 0,\phi}}\right)
&=& 0\,.\nonumber
\eea
In order to eliminate the pressure term, we calculate the $r$ derivative of the poloidal momentum equation (Equation \ref{eq_pert_ene_mom}) multiplied by a factor $(ir/kw_0)$ to $\mathcal{O}(v_1)$:
\bea
\label{eq_dFz_dr}
-\frac{\partial_{\ln r}p_1}{w_0}
+\left[\frac{l^2}{k^2}(1+\sigma_0)\tilde\omega+\frac{il}{k}\sigma_0u\right]iv_{\rm 1,r}
\nonumber\\
-\alpha_{\rm\phi}\sigma_0\frac{B_{\rm 1,\phi}}{B_{\rm 0,\phi}}
-\sigma_0\left(\frac{\partial_{\ln r}B_{\rm 1,\phi}}{B_{\rm 0,\phi}}-\frac{m}{kr}\frac{\partial_{\ln r}B_{\rm 1,z}}{B_{\rm 0,\phi}}\right)
&=& 0\,.
\eea
Summing Equation (\ref{eq_dFz_dr}) with Equation (\ref{eq_Fr}), we obtain:
\bea
\label{eq_Fr2}
\left(1+\frac{l^2}{k^2}\right)(1+\sigma_0)i\tilde\omega v_{\rm 1,r}
-i(m+\eta)\sigma_0\frac{B_{\rm 1,r}}{B_{\rm 0,\phi}}
\\
+2\sigma_0\frac{B_{\rm 1,\phi}}{B_{\rm 0,\phi}}
+\frac{(m+\eta)}{kr}\sigma_0\frac{\partial_{\ln r}B_{\rm 1,z}}{B_{\rm 0,\phi}}
&=& 0\,.\nonumber
\eea
We need to calculate $\partial_{\ln r}B_{\rm 1,z}$, but only to $\mathcal{O}(krv_1 B_{\rm 0,\phi})$:
\bea
\label{eq_dB1z_dr}
\frac{\partial_{\ln r}B_{\rm 1,z}}{B_{\rm 0,\phi}} &=&
kr(m+\eta)\left[-\frac{l^2}{k^2}+2\left(1-\frac{il}{k}\frac{\tilde\omega}{u}\right)\frac{u^2}{\tilde\omega^2}\right]\frac{iv_{\rm 1,r}}{\tilde\omega}.
\eea
After substituting the perturbed magnetic field to Equation (\ref{eq_Fr2}), we obtain:
\bea
\label{eq_Fr3}
\left(1+\frac{l^2}{k^2}\right)\left[(1+\sigma_0)\tilde\omega^2-(m+\eta)^2\sigma_0\right]\frac{iv_{\rm 1,r}}{\tilde\omega}
\\
+2\sigma_0\left[\alpha_{\rm\phi}+(m+\eta)^2\left(1-\frac{il}{k}\frac{\tilde\omega}{u}\right)\frac{u^2}{\tilde\omega^2}\right]\frac{iv_{\rm 1,r}}{\tilde\omega}
\nonumber\\
+2\sigma_0\frac{(\eta v_{\rm 1,\phi}-v_{\rm 1,lrkz})}{\tilde\omega}
&=& 0\,.
\nonumber
\eea
In the last term, we substitute the perturbation velocities using Equations (\ref{eq_v1phi}) and (\ref{eq_v1lrkz}). Finally, dividing Equation (\ref{eq_Fr3}) by a factor $(2i\sigma_0v_{\rm 1,r}/\tilde\omega)$, we obtain the dispersion relation:
\bea
\label{eq_disp}
\frac{1}{2}\left(1+\frac{l^2}{k^2}\right)\left[\left(1+\frac{1}{\sigma_0}\right)\tilde\omega^2-(m+\eta)^2\right]
\nonumber\\
+\alpha_{\rm\phi}+(m+\eta)^2\left(1-\frac{il}{k}\frac{\tilde\omega}{u}\right)\frac{u^2}{\tilde\omega^2}
\nonumber\\
+\frac{(m+\eta)^2v_{\rm s}^2\sigma_0-\tilde\omega^2(\sigma_0-v_{\rm s}^2)}{(m+\eta)^2v_{\rm s}^2\sigma_0-\tilde\omega^2(\sigma_0+v_{\rm s}^2)}
&=& 0
\eea
(see Appendix \ref{sec_app2} for a notation look-up table). For $u=0$, the above equation corresponds exactly to Equation (3.32) of \cite{1998ApJ...493..291B}. The velocity shear term can be traced back to the extra perturbed poloidal magnetic field $B_{\rm 1,z}$ (see Equations \ref{eq_B1z} and \ref{eq_dB1z_dr}) via the $\bm{j}_1\times\bm{B}_0$ force (Equation \ref{eq_j1xB0}). The $(m+\eta)^2$ factor indicates that it contributes only for $m\ne 0$ in the presence of a purely toroidal field ($\eta=0$), or requires a finite background poloidal magnetic field ($\eta\ne 0$). The velocity shear terms become dominant for $\omega\to 0$.

The KHI modes, driven by gas pressure in the presence of velocity shear, are suppressed in the short wavelength approximation. This can be seen by noting that the $\bm{v}_0\partial_tp_1$ force would contribute a term $-\omega up_1/kw_0$ to Equation (\ref{eq_dFz_dr}), which is of order $\mathcal{O}(v_1/kr)$. Even in the absence of the magnetic fields, it would be dominated by the acceleration term by one order in $kr$. The KHI modes would be important only when $\left|u\right|\gtrsim\mathcal{O}(kr)$.

\section{Stability analysis}
\label{sec_stab}

In the general case, the dispersion relation derived in the previous section (Equation \ref{eq_disp}) is equivalent to a sixth-order polynomial over the complex space of $\tilde\omega$. Hence, we cannot perform a fully analytic stability analysis, as is possible for the case of a static column, when Equation (\ref{eq_disp}) reduces to a quadratic equation for $\tilde\omega^2$ with only real solutions \citep{1998ApJ...493..291B}. Substituting $\tilde\omega=\tilde\omega_{\rm R}-i\tilde\omega_{\rm I}$, where $\tilde\omega_{\rm R},\tilde\omega_{\rm I}\in\mathcal{R}$, we can separate Equation (\ref{eq_disp}) into its real and imaginary components (we also substitute $a=(m+\eta)^2$ and $d=l/k$, see Appendix \ref{sec_app2}):
\bea
\label{eq_disp_real}
\frac{(1+d^2)}{2}\left[\left(1+\frac{1}{\sigma_0}\right)(\tilde\omega_{\rm R}^2-\tilde\omega_{\rm I}^2)-a\right]
+\alpha_{\rm\phi}
\nonumber\\
+au^2\frac{(\tilde\omega_{\rm R}^2-\tilde\omega_{\rm I}^2)}{(\tilde\omega_{\rm R}^2+\tilde\omega_{\rm I}^2)^2}
+\frac{adu\tilde\omega_{\rm I}}{(\tilde\omega_{\rm R}^2+\tilde\omega_{\rm I}^2)}
\\
+\frac{[av_{\rm s}^2-(\tilde\omega_{\rm R}^2-\tilde\omega_{\rm I}^2)]^2\sigma_0^2-(\tilde\omega_{\rm R}^2+\tilde\omega_{\rm I}^2)^2v_{\rm s}^4+4\tilde\omega_{\rm R}^2\tilde\omega_{\rm I}^2\sigma_0^2}
{[av_{\rm s}^2\sigma_0-(\tilde\omega_{\rm R}^2-\tilde\omega_{\rm I}^2)(\sigma_0+v_{\rm s}^2)]^2+4\tilde\omega_{\rm R}^2\tilde\omega_{\rm I}^2(\sigma_0+v_{\rm s}^2)^2}
&=& 0\,,
\nonumber\\
\label{eq_disp_imag}
(1+d^2)\left(1+\frac{1}{\sigma_0}\right)\tilde\omega_{\rm R}\tilde\omega_{\rm I}
-\frac{2au^2\tilde\omega_{\rm R}\tilde\omega_{\rm I}}{(\tilde\omega_{\rm R}^2+\tilde\omega_{\rm I}^2)^2}
+\frac{adu\tilde\omega_{\rm R}}{(\tilde\omega_{\rm R}^2+\tilde\omega_{\rm I}^2)}
\\
+\frac{4av_{\rm s}^4\sigma_0\tilde\omega_{\rm R}\tilde\omega_{\rm I}}{[av_{\rm s}^2\sigma_0-(\tilde\omega_{\rm R}^2-\tilde\omega_{\rm I}^2)(\sigma_0+v_{\rm s}^2)]^2+4\tilde\omega_{\rm R}^2\tilde\omega_{\rm I}^2(\sigma_0+v_{\rm s}^2)^2}
&=& 0\,,
\nonumber
\eea
respectively. The imaginary equation has a trivial solution at $\tilde\omega_{\rm R}=0$. Hence, for purely exponential solutions it is sufficient to solve only Equation (\ref{eq_disp_real}). We further note that Equation (\ref{eq_disp_real}), and Equation (\ref{eq_disp_imag}) divided by $\tilde\omega_{\rm R}$, contain only quadratic terms in $\tilde\omega_{\rm R}$. Hence, it is sufficient to look for solutions with $\tilde\omega_{\rm R}>0$. Also, we are interested solely in solutions with $\tilde\omega_{\rm I}>0$. In the case of vanishing velocity shear, the left-hand side of Equation (\ref{eq_disp_imag}) is always positive, hence there are no solutions with $\tilde\omega_{\rm R}>0$.

\begin{figure}
\includegraphics[width=\columnwidth]{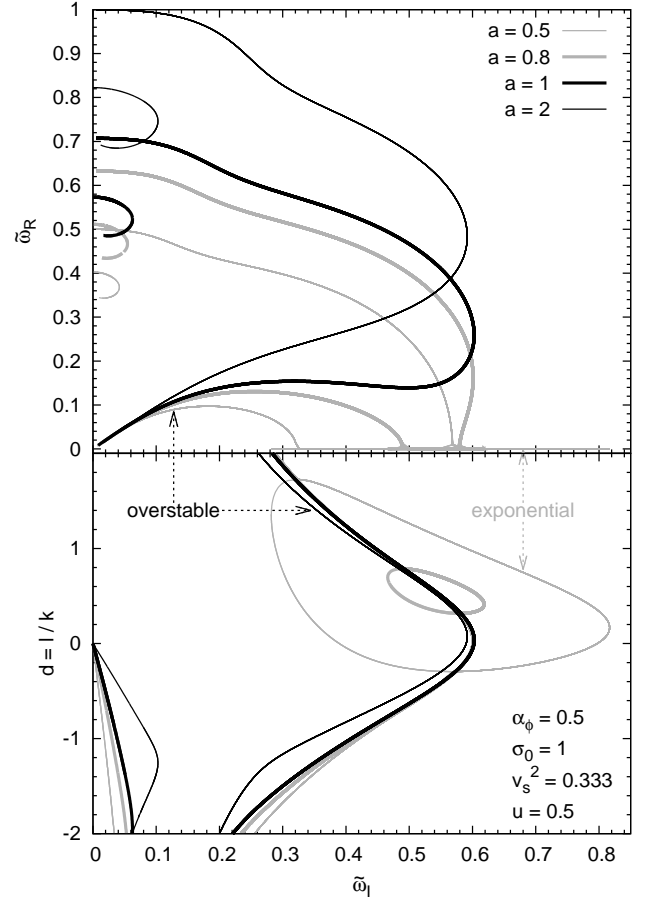}
\caption{Sample solutions to the system of dispersion relations given by Equations (\ref{eq_disp_real}) and (\ref{eq_disp_imag}). For each value of $a=(m+\eta)^2$, we plot solutions for a wide range of $d=l/k$. Both panels show the same set of solutions in different projections.}
\label{fig_disp1_mix1}
\end{figure}

In Figure \ref{fig_disp1_mix1}, we show typical solutions to the system of Equations (\ref{eq_disp_real}) and (\ref{eq_disp_imag}). We fix the values for $\alpha_{\rm\phi}$, $u$, $\sigma_0$ and $v_{\rm s}$, and for selected values of $a$ we continuously vary parameter $d$. For all values of $a$ and $d$, there are some overstable modes (with $\tilde\omega_{\rm R}\ne 0$), which form two separate branches in the $(\tilde\omega_{\rm R},\tilde\omega_{\rm I})$ plane. However, for sufficiently small values of $a$, there exists a class of exponential solutions, which dominate the overstable solutions in terms of the growth rate $\tilde\omega_{\rm I}$.

\begin{figure}
\includegraphics[width=\columnwidth]{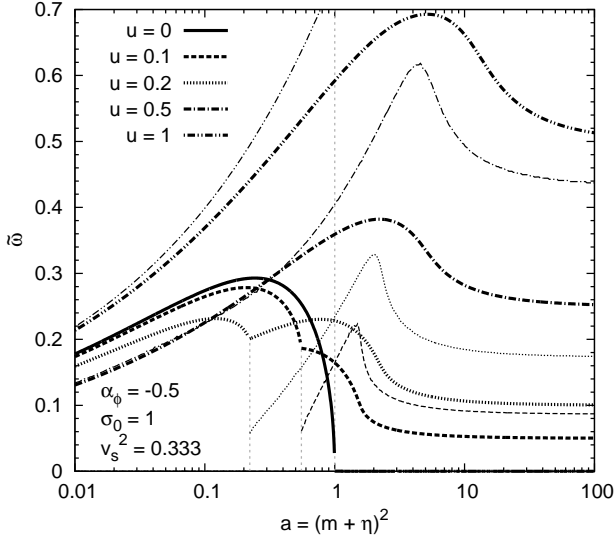}
\caption{Fastest-growing modes as functions of the parameter $a$, and for different values of the velocity shear $u$, obtained by varying the parameter $d$. \emph{Thick lines} show the growth rate $\tilde\omega_{\rm I}$, while \emph{thin lines} show the oscillation frequency $\tilde\omega_{\rm R}$. The \emph{dashed vertical} lines mark the transition between the exponential and overstable modes, with overstable modes dominating to the right of the line.}
\label{fig_disp1_mix2}
\end{figure}

In Figure \ref{fig_disp1_mix2}, we show the fastest-growing modes as a function of $a$, selected by varying $d$, for a fixed value of $\alpha_{\rm\phi}=-0.5$ and several values of the velocity shear $u$. For $u=0$, such a system is known to be unstable only for modes with $a<1$, which is consistent with the stability criterion $a<2(\alpha_{\rm\phi}+1)$ \citep[see Equation (4.2) in][]{1998ApJ...493..291B}. These unstable modes are purely exponential. For small non-zero values of $u$, overstable modes arise for $a>a_1$, where $a_1<1$. These modes are slower than the exponential ones for $u=0.1$, but of comparable growth rate for $u=0.2$. For $u\gtrsim 0.2$, the overstable modes dominate the exponential solutions, and their growth rate is roughly proportional to $u$. The value of $a$ corresponding to the fastest-growing modes increases with increasing $u$ for the overstable solutions, and decreases with increasing $u$ for the exponential solutions. The overstable solutions generally have their oscillatory frequency $\tilde\omega_{\rm R}>\tilde\omega_{\rm I}$.

\begin{figure}
\includegraphics[width=\columnwidth]{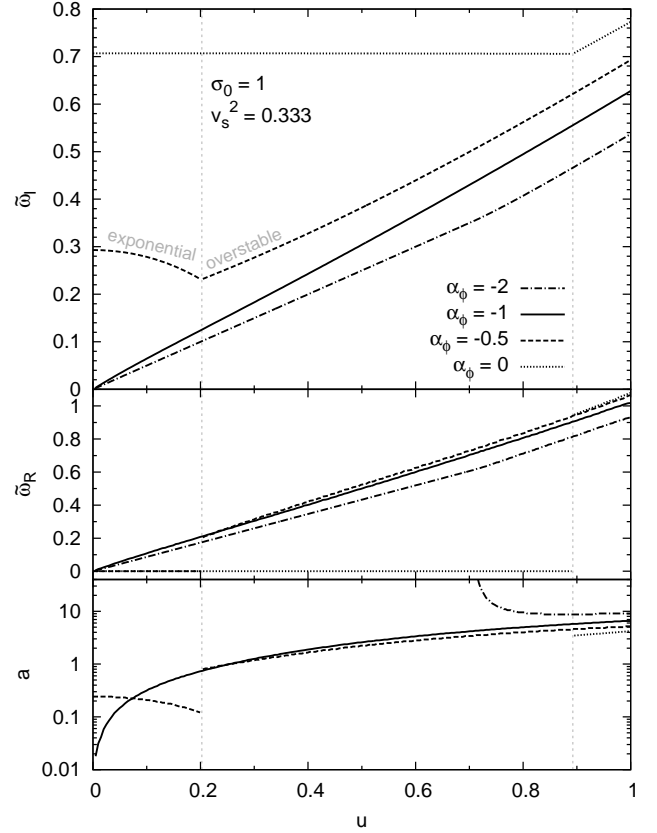}
\caption{Fastest-growing modes selected by varying parameters $d$ and $a$, as functions of the velocity shear $u$ for different values of parameter $\alpha_{\rm\phi}$. \emph{Top panel}: the growth rate $\tilde\omega_{\rm I}$. \emph{Middle panel}: the oscillation frequency $\tilde\omega_{\rm R}$. \emph{Bottom panel}: the optimal value of parameter $a$. For the cases having exponential solutions, the \emph{dashed vertical} lines mark the transition between the exponential and overstable modes.}
\label{fig_disp1_mix3}
\end{figure}

In Figure \ref{fig_disp1_mix3} we show the fastest-growing modes selected by varying both $d$ and $a$, as a function of the velocity shear $u$. In the case of $\alpha_{\rm\phi}=-0.5$, it is confirmed that the fastest-growing modes are purely exponential for $u\lesssim 0.2$ and overstable for $u\gtrsim 0.2$. We confirm the behavior of $a$ with $u$ identified in Figure \ref{fig_disp1_mix2}. In the case of $\alpha_{\rm\phi}=0$, the exponential modes with $a=d=0$ strongly dominate over the overstable modes and the velocity shear has very little impact on the former. The case of $\alpha_{\rm\phi}=-1$ corresponds to marginal stability in the absence of velocity shear. In this case, there are no purely exponential solutions, but the system is prone to overstable modes. Both the growth rate and the oscillation rate are proportional to the velocity shear, with approximate scaling relations $\tilde\omega_{\rm I}\simeq 0.62u$ and $\tilde\omega_{\rm R}\simeq u$. Overstable modes are present also for $\alpha_{\rm\phi}=-2$.

\begin{figure}
\includegraphics[width=\columnwidth]{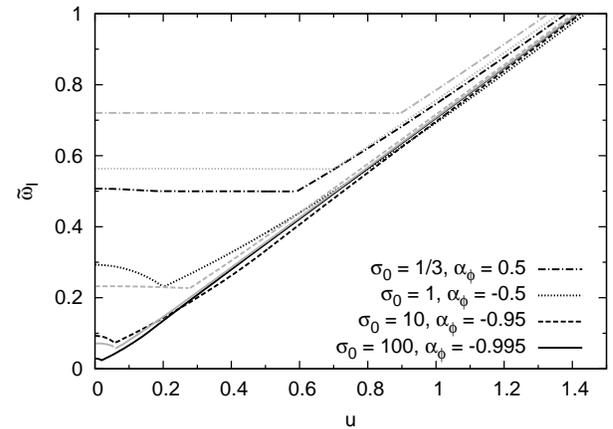}
\caption{Dependence of the fastest growth rates on the plasma magnetization $\sigma_0$ and the sound speed $v_{\rm s}$. The toroidal magnetic field distribution is chosen to satisfy $(\alpha_{\rm\phi}+1)=1/(2\sigma_0)$. The \emph{black lines} show the results for $v_{\rm s}^2=1/3$, and the \emph{gray lines} show the results for $v_{\rm s}^2=0.1$.}
\label{fig_disp1_mix4}
\end{figure}

Figure \ref{fig_disp1_mix4} shows the dependence of the fastest growth rates on the magnetization $\sigma_0$ and the sound speed $v_{\rm s}$. When changing the plasma magnetization, we need to adjust the toroidal magnetic field distribution $\alpha_{\rm\phi}$ to maintain a realistic radial force balance. We use Equation (\ref{eq_fr}), setting $v_{\rm 0,z}=0$ and $p_0=w_0/4$ to obtain a relation $(\alpha_{\rm\phi}+1)=-\zeta/(4\sigma_0)$. For $\zeta=0$, i.e. a uniform distribution of the gas pressure, we have $\alpha_{\rm\phi}=-1$. In Figure \ref{fig_disp1_mix4}, we show a non-trivial case of $\zeta=-2$. We find that the plasma magnetization has a significant effect on the growth rates of the exponential modes and the minimum velocity shear for which the overstable modes dominate the exponential modes, but it has a weak effect on the growth rates of the overstable modes. For a given value of the velocity shear $u$, the growth rate decreases with increasing $\sigma_0$, i.e., highly magnetized plasmas are more stable. The limit of $\sigma_0\gg 1$ is called the force-free limit. In this case, the dimensionless growth rate for the exponential modes tends to zero, and the overstable modes can dominate the exponential modes for very small values of $u$. For moderate values of $\sigma_0$, the growth rate is sensitive to the sound speed (i.e. to the plasma temperature), being higher for colder plasmas.

Our results indicate that the overstable modes can be driven for any value of $\alpha_{\rm\phi}$, $\sigma_0$ and $v_{\rm s}$, and that they dominate other solutions for a sufficiently large value of $u$. Both the growth rate of the overstable modes and the oscillation frequency are roughly proportional to the velocity shear.

\section{Discussion}
\label{sec_appl}

In order to evaluate the importance of the velocity shear in affecting the stability of astrophysical jets, we first need to estimate the magnitude of this parameter. Our calculations in previous sections were performed in the local co-moving frame (in this section, we denote the quantities calculated in the co-moving frame with a prime). The co-moving velocity shear is directly related to the transverse gradient of the Lorentz factor calculated in the external frame:
\be
\label{eq_u_Gamma}
u'=r\partial_rv_{\rm 0,z}'=\Gamma^2r\partial_rv_{\rm 0,z}=\frac{1}{v_{\rm 0,z}}\pder{\ln\Gamma}{\ln r}\,.
\ee
In relativistic jets, a co-moving velocity shear value of $u'=1$ corresponds to the doubling of the Lorentz factor (as measured by an outside observer) with a doubling of the radius. The co-moving velocity shear is significant when the Lorentz factor varies by a sizable factor across the jet.

\begin{figure}
\includegraphics[width=\columnwidth]{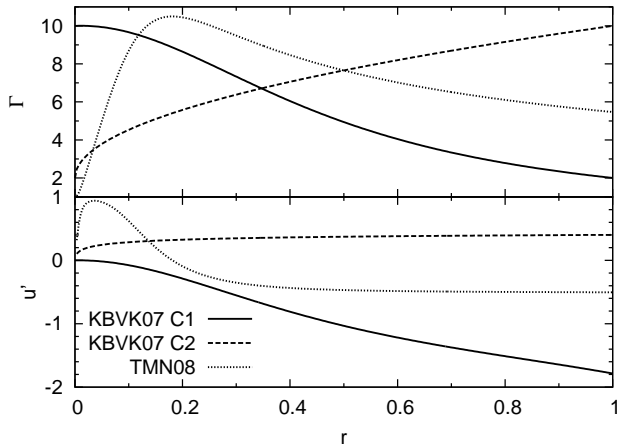}
\caption{\emph{Top panel:} three functions for the Lorentz factor $\Gamma$ distribution across a jet, motivated by numerical results of Komissarov et~al. (2007) and Tchekhovskoy et~al. (2008). \emph{Bottom panel:} the corresponding distribution of the co-moving velocity shear $u'$.}
\label{fig_toy1}
\end{figure}

Relativistic jets are thought to be accelerated by the action of magnetic stresses in an extended region dominated by the Poynting flux. Both analytical studies \citep{2004ApJ...605..656V,2009ApJ...698.1570L} and numerical simulations \citep{2007MNRAS.380...51K,2008MNRAS.388..551T,2010ApJ...709.1100P} show that this process does not operate with uniform efficiency across the jet radius. This results in a strong dependence of the Lorentz factor within the jet on the radial coordinate (equivalently on the polar angle). To illustrate these dependencies, in the top panel of Figure \ref{fig_toy1}, we plot three models for the radial variation of the Lorentz factor, qualitatively based on the results of \cite{2007MNRAS.380...51K} (models C1 and C2) and \cite{2008MNRAS.388..551T} (Appendix 4.1). In the bottom panel of Figure \ref{fig_toy1}, we plot the corresponding values of the co-moving velocity shear given by Equation (\ref{eq_u_Gamma}). We find that velocity shear values of order $\left|u'\right|\sim 0.5$ and higher can be expected in relativistic jets. This indicates that the effect of the poloidal velocity shear on the stability of jets is very important. A gradual jet acceleration also means that the importance of velocity shear can be expected to increase with distance.

In our local approximation, the sign of the velocity shear has no effect on the maximum growth rate for a given system; only the magnitude of the shear is important. This is because in Equation (\ref{eq_disp}), the linear term in $u$ is coupled with the $l/k$ ratio, and a change in the sign of $u$ can be countered by a change in the sign of $l$ (inward or outward propagation) or $k$ (upstream or downstream propagation). On the other hand, \cite{2009ApJ...697.1681N} found that for global modes in the force-free approximation, the sign of the poloidal velocity shear is important. They found that jets with positive velocity shear (Lorentz factor increasing with radius) are stable, while those in which the velocity shear changes sign at an intermediate radius, as in the example shown in our Figure \ref{fig_toy1} with a dotted line, are unstable. This fundamental difference in conclusions reached via global and local analysis is not surprising. It demonstrates that even in globally stable jets, vigorous dissipation may result from the presence of local modes.

The local stability criterion formulated by \cite{1998ApJ...493..291B} for a static plasma column, which can be written as $a>2(\alpha_{\rm\phi}+1)$, is no longer valid if poloidal velocity shear is present. This is clearly illustrated in Figure \ref{fig_disp1_mix2}. In the case of $\alpha_{\rm\phi}=-0.5$, the stability criterion is $a>1$. However, even for a very small value of $u$, there exist overstable modes with $a>1$. Even for $\alpha_{\rm\phi}\le -1$, in which case the static equilibria are always stable, we find overstable solutions.

A mode with a co-moving dimensionless growth rate of $\tilde\omega_{\rm I}'$ will double its amplitude over the co-moving time scale of $t_2'=0.7r/\tilde\omega_{\rm I}'$. Over this time, an element located initially at $z$ will travel a distance $z_2$ in the external frame given by $(z_2/z)\sim 0.7(\Gamma\theta)/\tilde\omega_{\rm I}'$, where $\Gamma\gg 1$ is the local Lorentz factor and $\theta=r/z$ is the polar angle. Since in relativistic jets we expect $\Gamma\theta\lesssim 1$, for $\tilde\omega_{\rm I}'\sim\mathcal{O}(1)$ the growth length measured in the external frame is not particularly short, with $(z_2/z)\sim\mathcal{O}(1)$, with the exception of the core region close to the jet axis. For a typical co-moving growth rate of $\tilde\omega_{\rm I}'\sim 0.4$, the energy dissipation associated with the local CDI can be expected to proceed over at least one order of magnitude in distance along the jet. A major caveat is that the growth of local CDI modes competes with lateral jet expansion. The question of whether jet expansion is able to quench the CDI modes must be addressed by a careful analysis of realistic jet models.

The results of this work need to be confirmed by numerical simulations. We are currently preparing local simulations of cylindrical plasma columns similar to those in \cite{2012MNRAS.422.1436O}, adding a background poloidal velocity shear. However, this issue also needs to be addressed by global simulations, like those in \cite{2009MNRAS.394L.126M}, but with a much higher resolution.

Our study of the local development of CDI is still very simplified, as we do not take into account several effects that could potentially have an impact on the CDI growth rate. Some of these effects, like rotation or lateral expansion, are not supposed to be important at the scales we are interested in. However, centrifugal forces due to jet collimation, enhanced by relativistic effects, can significantly change the radial force balance \citep{2009ApJ...697.1681N} and should be included in future work.

\section{Summary}
\label{sec_summ}

We have calculated the local dispersion relation (Equation \ref{eq_disp}) for a cylindrical plasma column with smooth poloidal velocity shear (defined as $u=r\partial_rv_{\rm 0,z}$), extending the original results of \cite{1998ApJ...493..291B}. The dispersion relation is a sixth-order complex equation for dimensionless frequency $\tilde\omega=\omega r=\tilde\omega_{\rm R}-i\tilde\omega_{\rm I}$, which allows two types of solutions: `exponential' solutions, which have a purely imaginary frequency $\tilde\omega_{\rm I}>0$; and `overstable' solutions that in addition have a non-zero real frequency $\tilde\omega_{\rm R}\ne 0$. We explore the parameter space of mode wavenumbers and identify the fastest-growing modes for a given equilibrium. The growth rate of the exponential modes decreases with increasing velocity shear, while that of the overstable modes is roughly proportional to the shear. For a sufficiently high shear the overstable modes dominate the exponential modes. The transition between the two regimes depends on the distribution of the toroidal magnetic field (parametrized by $\alpha_{\rm\phi}=\partial_{\ln r}(\ln B_{\rm 0,\phi})$), the plasma magnetization ($\sigma_0=B_0^2/(4\pi w_0)$) and the plasma temperature (measured by the sound speed given by $v_{\rm s}^2=\gamma p_0/w_0$). In the force-free limit ($\sigma_0\gg 1$), as $\alpha_{\rm\phi}\simeq -1$, the exponential modes become unimportant, and the overstable modes dominate over a wide range of shear values.

These results can be applied to relativistic jets of AGN and GRBs, in the initial force-free region, in the main acceleration region, and beyond. Studies of magnetic acceleration of jets predict a strong dependence of the jet Lorentz factor on the cylindrical radius, which translates to significant values of the co-moving velocity shear. Hence, in many realistic jet configurations, the overstable modes identified in this work may dominate locally. Since these modes are present for every set of equilibrium parameters, they are very hard to suppress unless the underlying velocity shear is reduced to insignificant values. Local CDI may play an important role in regulating jet acceleration and collimation. In the meantime, a significant fraction of the relative kinetic energy may be dissipated, transferred into ultra-relativistic particles and converted into non-thermal radiation that dominates the spectra of blazars and GRBs.

\section*{Acknowledgments}
We thank Sean O'Neill for his comments on the manuscript. This work was partly supported by NSF grant AST-0907872 and NASA ATP grant NNX09AG02G.

\appendix

\section{Individual terms of the perturbed energy-momentum equation}
\label{sec_app1}

Here, we calculate individual terms of Equation (\ref{eq_pert_ene_mom}). We retain the terms linear in $v_{\rm 0,z}$ only in the $z$ component, as it will be differentiated with respect to $r$. The acceleration term is:
\bea
\Gamma_0^2w_0\left[\partial_t\bm{v}_1+(\bm{v}_1\cdot\bm\nabla)\bm{v}_0+(\bm{v}_0\cdot\bm\nabla)\bm{v}_1\right]
&\simeq&
\nonumber\\
\simeq
i\omega w_0\bm{v}_{\rm 1}
+\frac{w_0}{r}\left(ikrv_{\rm 0,z}v_{\rm 1,z}+uv_{\rm 1,r}\right)\hat{\bm z}\,.
\eea
Gas pressure forces are:
\bea
\bm{v}_0\partial_tp_1 &=&
i\omega v_{\rm 0,z}p_1\hat{\bm z}\,,
\\
\bm\nabla p_1 &=& \partial_rp_1\hat{\bm r}+\frac{imp_1}{r}\hat{\bm\phi}+ikp_1\hat{\bm z}\,.
\eea
Electromagnetic forces are:
\bea
-\rho_{\rm e,1}\bm{E}_0 &=& 0\,,
\\
-\rho_{\rm e,0}\bm{E}_1
&=&
\frac{B_{\rm 0,\phi}}{4\pi r}u(B_{\rm 0,z}v_{\rm 1,\phi}-B_{\rm 0,\phi}v_{\rm 1,z})\hat{\bm r}
\nonumber\\
&&
-\frac{B_{\rm 0,\phi}B_{\rm 0,z}}{4\pi r}uv_{\rm 1,r}\hat{\bm\phi}
\nonumber\\
&&
+\frac{B_{\rm 0,\phi}^2}{4\pi r}\left[(1+\alpha_{\rm\phi})v_{\rm 0,z}+u\right]v_{\rm 1,r}\hat{\bm z}\,,
\\
\label{eq_j1xB0}
-\bm{j}_1\times\bm{B}_0
&=&
i\sigma_0\omega w_0v_{\rm 1,r}\hat{\bm r}
\nonumber\\
&&
+\frac{B_{\rm 0,\phi}}{4\pi r}\left[B_{\rm 1,\phi}-i(m+\eta)B_{\rm 1,r}\right]\hat{\bm r}
\nonumber\\
&&
+\frac{1}{4\pi}\left(B_{\rm 0,\phi}\partial_rB_{\rm 1,\phi}+B_{\rm 0,z}\partial_rB_{\rm 1,z}\right)\hat{\bm r}
\nonumber\\
&&
+\frac{B_{\rm 0,z}}{4\pi}i\omega(v_{\rm 1,\phi}B_{\rm 0,z}-v_{\rm 1,z}B_{\rm 0,\phi})\hat{\bm\phi}
\nonumber\\
&&
+\frac{i}{4\pi r}(mB_{\rm 1,z}B_{\rm 0,z}-\eta B_{\rm 1,\phi}B_{\rm 0,\phi})\hat{\bm\phi}
\nonumber\\
&&
+\frac{B_{\rm 0,\phi}}{4\pi}i\omega(v_{\rm 1,z}B_{\rm 0,\phi}-v_{\rm 1,\phi}B_{\rm 0,z})\hat{\bm z}
\nonumber\\
&&
+\frac{iB_{\rm 0,\phi}^2}{4\pi r}\left(i\alpha_{\rm\phi}v_{\rm 1,r}+\eta v_{\rm 1,\phi}-v_{\rm 1,lrkz}\right)v_{\rm 0,z}\hat{\bm z}
\nonumber\\
&&
+\frac{B_{\rm 0,\phi}}{4\pi r}\left(ikrB_{\rm 1,\phi}-imB_{\rm 1,z}\right)\hat{\bm z}\,,
\\
-\bm{j}_0\times\bm{B}_1 &=& 
(\alpha_{\rm\phi}+1)\frac{B_{\rm 0,\phi}}{4\pi r}\left(B_{\rm 1,\phi}\hat{\bm r}-B_{\rm 1,r}\hat{\bm\phi}\right)\,,
\\
(\bm{E}_1\cdot\bm{j}_0)\bm{v}_0 &=&
-(\alpha_{\rm\phi}+1)\frac{B_{\rm 0,\phi}^2}{4\pi r}v_{\rm 0,z}v_{\rm 1,r}\hat{\bm z}\,,
\\
(\bm{E}_0\cdot\bm{j}_1)\bm{v}_0 &=& 0\,.
\eea

\clearpage

\section{Notation}
\label{sec_app2}

\begin{tabular}{cl}
$\omega=\omega_{\rm R}-i\omega_{\rm I}$ & mode frequency \\
$\tilde\omega=\omega r$ \\
$l,m,k$ & wave numbers \\
$d = l / k$ \\
$B_{\rm 0,\phi},B_{\rm 0,z}$ & magnetic field \\
$\alpha_{\rm\phi}=\partial_{\ln r}(\ln B_{\rm 0,\phi})$ \\
$\eta=kr(B_{\rm 0,z}/B_{\rm 0,\phi})$ \\
$a = (m+\eta)^2$ \\
$w_0$ & specific enthalpy \\
$\sigma_0=B_0^2/(4\pi w_0)$ & magnetization \\
$p_0$ & gas pressure \\
$\zeta = \partial_{\ln r}(\ln p_0)$ \\
$\gamma$ & adiabatic index \\
$v_{\rm s}^2=\gamma p_0/w_0$ & sound speed \\
$v_{\rm 0,z}$ & poloidal velocity \\
$u=\partial_{\ln r}v_{\rm 0,z}$ & velocity shear
\end{tabular}

\end{document}